\documentclass[prb,twocolumn,showpacs]{revtex4}
\usepackage{graphicx}
\usepackage{amsmath}
\usepackage{amssymb}
\usepackage{epstopdf}
\usepackage{slashed}

\begin{document}
\title{Fractional quantum Hall states in two-dimensional electron systems with anisotropic interactions}
\author{Hao Wang,$^1$ Rajesh Narayanan,$^{1,2,3}$ Xin Wan,$^4$ and Fuchun Zhang$^{1,4}$}
\affiliation{$^1$Department of Physics, The University of Hong Kong,
Hong Kong SAR, China \\ $^2$Department of Physics, Indian Institute of Technology Madras, Chennai 600036, India\\
$^3$Department of Physics, Hong Kong University of Science and Technology, Hong Kong SAR, China\\ $^4$Zhejiang Institute of
Modern Physics, Zhejiang University, Hangzhou 310027, China }

\begin{abstract}
We study the anisotropic effect of the Coulomb interaction on a
1/3-filling fractional  quantum Hall system by using exact
diagonalization method on small systems in torus geometry. For weak
anisotropy the system remains to be an incompressible quantum
liquid, although anisotropy manifests itself in density correlation
functions and excitation spectra. When the strength
of anisotropy increases, we find the system
develops a Hall-smectic-like phase with one-dimensional charge density wave order and is unstable towards the one-dimensional crystal in the strong anisotropy limit. In all three
phases of the Laughlin liquid, Hall-smectic-like, and crystal phases the
ground state of the anisotropic Coulomb system can be well described
by a family of model wavefunctions generated by an anisotropic
projection Hamiltonian. We discuss the relevance of the results to
the geometrical description of fractional quantum Hall states
proposed by Haldane [Phys. Rev. Lett. {\bf 107}, 116801 (2011)].
\end{abstract}

%73.43.Cd    Theory and modeling
%73.43.Nq    Quantum phase transitions
%71.10.Pm   Fermions in reduced dimensions (anyons, composite fermions, Luttinger liquid, etc.)
%73.43.-f    Quantum Hall effects
\maketitle

\section{Introduction}

The fractional quantum Hall (FQH) effect at an odd denominator filling of $\nu$
has been understood as a property of incompressible quantum liquid
in an interacting two-dimensional (2D) electron system. Laughlin's
trial wavefunction \cite{laughlin:1983} is the first successful
theory to describe this many-body effect, where the interacting
system is implicitly assumed to be isotropic. Since then most
theoretical works on the FQH system have followed this simple
assumption and the FQH states are considered to be isotropic with
rotational symmetry.

However, the real FQH systems may be anisotropic. One natural source
for this is the anisotropic dielectric tensor, which in turn leads to
anisotropic Coulomb interaction. Other
mechanisms for various anisotropic FQH systems have also been
discussed theoretically
\cite{haldane:2011,qiu:2011,mulligan:2010,qiu:2012,haldane:2011b}
and experimentally.\cite{ni:2008,xia:2011} For example, an
anisotropic FQH state in a $\nu=7/3$ system has been observed in
experiment.\cite{xia:2011} In these anisotropic FQH systems,
rotational symmetry of the consequent ground state is expected to be
broken. In the extreme anisotropic interaction limit, where the
Coulomb interaction can be effectively treated to be one-dimensional
(1D), the ground state of the system will be a quasi-1D
crystal.\cite{aizenman:2010} Thus, the properties of the FQH state
in an anisotropic interaction may not always be associated with the
isotropic incompressible liquid. A comprehensive investigation on
the effect of the interaction anisotropy is called for.

Motivated by the anisotropic transport properties experimentally
reported at the partially filled higher Landau level (LL), trial
wavefunctions
\cite{balents:1996,musaelian:1996,ciftja:2001,fogler:2004} have been
proposed to describe the anisotropic FQH states. These variational
wavefunctions modify the original isotropic Laughlin wavefunction by
splitting the multiple-order zeros in the wavefunction. Very
recently, Haldane \cite{haldane:2011} has constructed a family of
the Laughlin states, which are the exact ground states of the
corresponding projected Hamiltonians and can be parameterized
according to the interaction anisotropy. This variational state can
be compared numerically with the anisotropic FQH state. An effort to map the underlying wavefunction of this
variational state has been reported.\cite{qiu:2012} In this paper we
study FQH states in 2D electron systems with
anisotropic Coulomb interaction and discuss the relevance of our
results with the geometric description of the FQH
states.\cite{haldane:2011}

The paper is organized as follows.  In Section~\ref{model} we
introduce our model with anisotropic Coulomb interaction and set up
the Hamiltonian on a torus geometry. In Section~\ref{result} we
discuss the properties of $\nu=1/3$ FQH states at different regimes
of the interaction anisotropy using energy spectra, charge density
and correlation functions. We also compare the anisotropic FQH state
with variational Laughlin states using wavefunction overlap.
Section~\ref{summary} summarizes the paper.

\section{Model and Numerical Setup}
\label{model}

We study a 2D electron system under a perpendicular magnetic field
$\mathbf{B}=B\hat{z}$. The electron-electron Coulomb interaction
with an in-plane biaxial dielectric tensor has the form
\begin{eqnarray}
V_{c}(\mathbf{r})&=&\frac{e^2}{4\pi\varepsilon\sqrt{A_{c}x^2+y^2/A_{c}}},
\label{potential}
\end{eqnarray}
where $A_c$ is the interaction anisotropy parameter and directions
of $\hat{x}$ and $\hat{y}$ are along the two principal axes of the
dielectric tensor. The effective mass tensor is considered isotropic
so that non-interacting electrons move in the circular cyclotron
orbitals. However, equipotential lines of the Coulomb interaction
are generally elliptical with $A_{c}\neq 1$. In the following
discussion, we choose $A_c\geq 1$ such that $\hat{x}$ is the hard
axis. For $A_c < 1$, one simply swaps the easy and hard axes. At
$A_c \gg 1$, the Coulomb interaction is effectively a 1D repulsion
along the hard axis.

In our numerical calculations, we use Landau gauge $(0,Bx)$ for the magnetic vector potential. Periodic boundary conditions for the magnetic translational operators are imposed with a quantized flux
$N_{\phi}$ through the rectangular unit cell $\mathbf{L_x} \times
\mathbf{L_y}$. The magnetic length $\ell$ is taken as the unit
length and the energy is in units of $e^2/4\pi \varepsilon \ell$. To
reduce the size of the Hilbert space, we carry out our calculation
at every pseudomomentum $\mathbf{K}=(K_x,K_y)$,\cite{rezayi:2000}
where $K_x$ ($K_y$) is in units of $2\pi/L_x$ ($2\pi/L_y$). The magnetic
field is assumed to be strong enough so that the spin degeneracy of
the Landau levels is lifted.\cite{morf:1998,rezayi:2000} One can
thus project the system Hamiltonian into the valence Landau
level.\cite{rezayi:2000} For the lowest Landau level, the projected
Hamiltonian has the form
\begin{eqnarray}
  H_{c}=
    \frac{1}{N_{\phi}} \sum_{\mathbf{q}} V(\mathbf{q}) e^{-q^2/2}
\sum_{i<j}e^{i \mathbf{q} \cdot (\mathbf{R}_i-\mathbf{R}_j)},
\label{hamiltonian}
\end{eqnarray}
where the momentum $\mathbf{q}=(q_x,q_y)$ takes discrete values
suitable for the lattice of the unit cell and $\mathbf{R}_i$ is the
guiding center coordinate of the $i$-th electron.
$V(\mathbf{q})=1/\sqrt{q_x^2/A_c+A_{c}q_y^2}$ is the Fourier
transform of the Coulomb interaction. From the geometrical point
of view, we generalize $q^2=q_x^2+q_y^2$ to $q_g^2=g^{ab} q_a q_b$,
where
\begin{equation}
\label{metric}
g = \left (
\begin{array}{cc}
1/A_c & 0 \\
0 & A_c
\end{array}
\right )
\end{equation}
is the inverse metric for the Coulomb interaction.

\section{Numerical Results and Discussion}\label{result}
For a $\nu=1/3$ FQH system, we first study the low-lying
energy spectra using exact diagonalization method. Here and in following
subsections, the default size of the $\nu=1/3$ system is $N_e=10$ and the
default shape of the unit cell is square unless otherwise specified. We find qualitatively similar results in systems with
other sizes and/or different shapes of the unit cell.

\begin{figure}[t]
\centerline{\includegraphics [width=3.4 in] {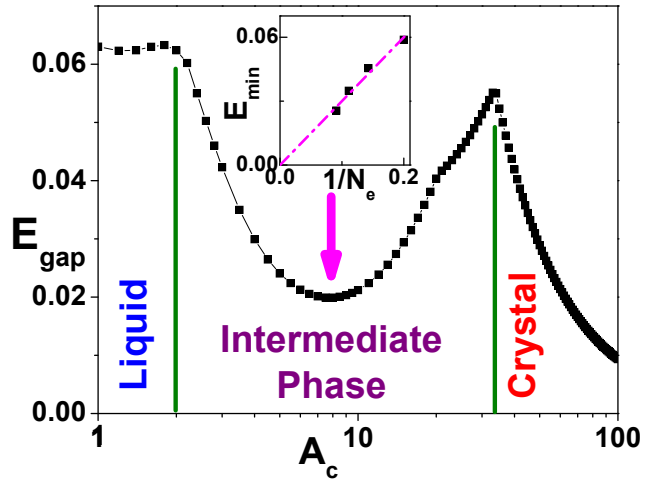}}
\caption{(Color online) Excitation energy gap versus interaction
anisotropy for $N_e=10$ and $\nu=1/3$ FQH system with a square unit
cell. The horizontal axis is plotted in the Log scale to show the
transition at low anisotropy. The system is in a Laughlin-liquid-like
state for $A_c < 2.0$ (Sec.~\ref{liquid}), and becomes a
quasi-1D crystal for $A_c > 33.0$ (Sec.~\ref{crystal}), whose
excitation gap scales as $1/\sqrt{A_c}$. The intermediate regime is
related to a Hall-smectic-like phase, which is unstable towards the quasi-1D crystal and will be discussed in
Sec.~\ref{transition}. The inset shows a linear size-scaling for the
minimum gap in the intermediate regime.} \label{Egap}
\end{figure}

Figure \ref{Egap} plots the excitation energy gap as a function  of
the Coulomb interaction anisotropy. In the range up to $A_c = 100$
the curve is nonmonotonic and develops several distinct regimes.
For small interaction anisotropy up to $A_c=2.0$, the energy gap
remains nearly constant, indicating the incompressible liquid phase
in the isotropic case (i.e., $A_c=1$) is robust against weak
interaction anisotropy. When the interaction anisotropy further
increases, the energy gap decreases to a minimum at around $A_c=8.0$.
For larger $A_c$ the energy gap increases with the interaction
anisotropy to a maximum at around $A_c =33.0$. The finite-size scaling
shown in the inset reveals that the minimum gap can close in
the thermodynamic limit, suggesting there might exist a switch
between different order parameters ruling the system. Beyond
$A_c=33.0$ the energy gap decreases roughly as $1/\sqrt{A_c}$,
indicating the regime of the quasi-1D repulsion limit. We have
studied other system sizes and found that these boundaries are
size-dependent. But, in general, the ground state of the system
maintains a three-fold degeneracy. This adiabatic transition with
complex regimes typically occurs between distinct phases with a
competition in the intermediate region. In the following
subsections, we will focus on these different regimes in $A_c$ and
reveal an interesting competition between liquid and crystal
phases.

\subsection{Anisotropic Laughlin Liquid at Small Interaction
Anisotropy}
\label{liquid}

\begin{figure}[t]
\centerline{\includegraphics [width=3.4 in] {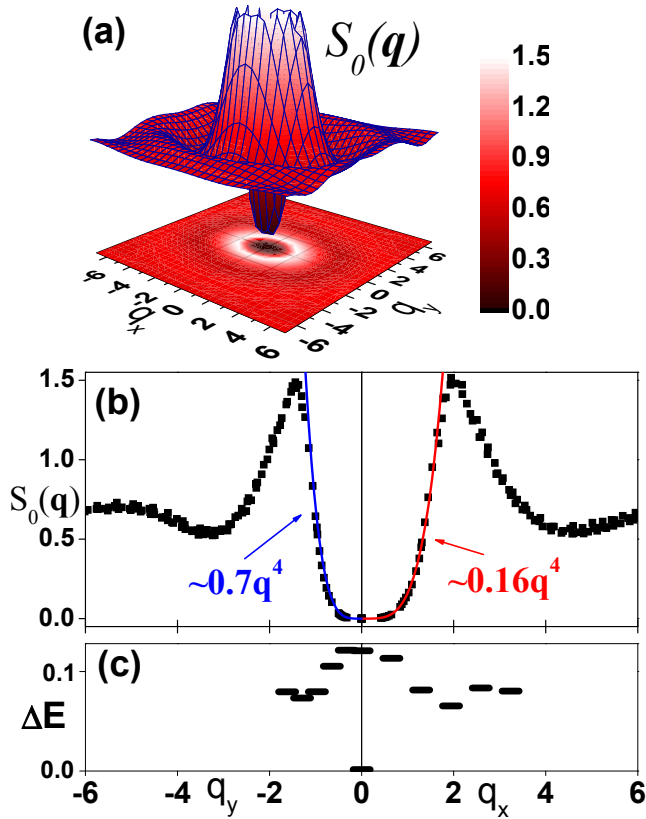}}
\caption{(Color online) (a) 3D and contour plots of the structure
factor, (b) structure factor along $q_x$ and $q_y$ axes, and (c)
excitation spectrum along $q_x$ and $q_y$ axes for the $N_e=10$
system with Coulomb interaction anisotropy $A_c=1.8$. To overcome
the discrete momentum limitation, we use unit cells with different
aspect ratios $R_a=L_x/L_y$ in (b) to obtain more data points and a
unit cell with $R_a=0.5$ in (c).} \label{sq}
\end{figure}

The energy gap plot suggests that the ground  state at small
interaction anisotropy is an incompressible liquid similar to the
isotropic Laughlin state. The anisotropy in interaction, however, is
expected to be imprinted, e.g., in the static structure factor of
the resulting incompressible liquid. The projected static structure
factor is defined as\cite{rezayi:2000}
\begin{eqnarray}
S_0(\mathbf{q})=\frac{1}{N_e}\langle 0|\sum_{i\neq
j}e^{i\mathbf{q}\cdot(\mathbf{r}_i-\mathbf{r}_j)}|0\rangle,
\label{strucfac}
\end{eqnarray}
where $|0\rangle$ is the calculated ground state and $\mathbf{r}_i$
is the coordinate of the $i$th particle.

In Fig. \ref{sq}(a), we draw the three-dimensional (3D) and contour
plots of the structure factor for a calculated FQH state at
$A_c=1.8$. It exhibits a crater-like feature, which is similar to
that of the isotropic liquid. However, the overall shape of the
crater is deformed, stretching along the hard axis direction.
Therefore, the elliptical symmetry replaces the circular symmetry in
the isotropic liquid case.

This anisotropic signature is more prominent in the 2D cuts along
the two principal axes as shown in Fig. \ref{sq}(b). We note that
the structure factor behaves asymptotically as $\sim q^4$ in the
long wavelength limit. This agrees with the single mode
approximation \cite{girvin:1985} (SMA) for incompressible liquid.
However, the prefactor of the quartic term is orientation dependent,
revealing the anisotropic nature of the structure factor. According
to Ref.~[\onlinecite{haldane:2011b}], the ratio of prefactors at
$q_x$ and $q_y$ axes is equal to $(1/A_L^*)^4$, where the parameter
$A_L^*$ defines an intrinsic metric, describing how the correlated
quasi-particles bind to each other in the anisotropic environment.
The fitting lines in the plot have revealed $A_L^*\sim 1.45$ for
interaction anisotropy $A_c=1.8$. The peaks in the
orientation-dependent plots represent the crater ridge in Fig.
\ref{sq}(a).

According to the SMA, the maximum in the structure factor
corresponds to a minimum gap in the excitation spectrum, or the
roton minimum, which corresponds to the excitonic binding of the
neutral quasiparticle-quasihole pairs.\cite{girvin:1985} This is
evident in Fig. \ref{sq}(c), where we plot the orientation-dependent
low-energy excitation spectra in the momentum space. The location of
the roton minimum is sensitive to the direction, but the gap value
is less sensitive. The ratio of the roton-minimum locations along
$q_x$ and $q_y$ axes is found close to $A_L^*$ as expected and these two locations
match the peak locations of the structure factor in Fig.
\ref{sq}(b).

\begin{figure}[t]
\centerline{\includegraphics [width=3.4 in] {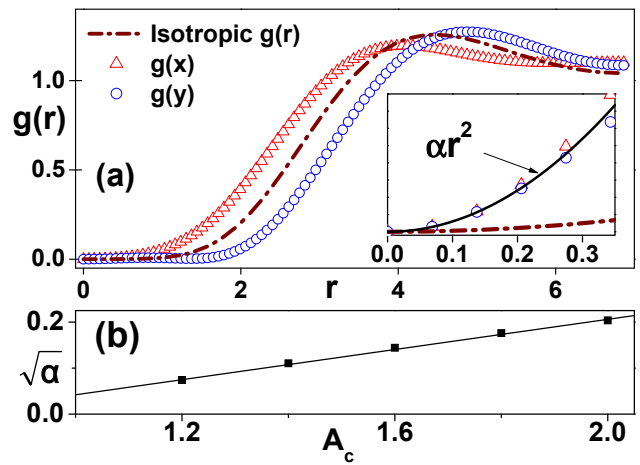}}
\caption{(Color online) (a) Pair correlation function along
$\hat{x}$ and $\hat{y}$ axes for the $N_e=10$ system at  interaction
anisotropy $A_c=1.8$. The dotted-dashed line represents the
correlation function at $A_c=1$ for comparison. The inset reveals
the $\alpha r^2$ behavior at small $r$ for the anisotropic case. (b)
A linear fit for the prefactor $\sqrt{\alpha}$ at small anisotropy.}
\label{gr}
\end{figure}

The isotropic Laughlin wavefunction, with order-3 zeros at the
locations  of other particles, triumphed in the explanation of the
isotropic incompressible liquids of $\nu=1/3$ FQH system.
Corresponding to the deformed electron-hole correlation from the anisotropic
interaction, order-3 zeros in the wavefunction are expected to
split. Several works have suggested that the relative coordinate
part of the anisotropic wavefunction
\cite{balents:1996,musaelian:1996,ciftja:2001,fogler:2004,qiu:2012}
has the form
\begin{eqnarray}
w({z_i}) \sim \prod_{i<j}(z_i-z_j)((z_i-z_j)^2 + z_0^2),
\end{eqnarray}
where $z_i$ is the complex coordinate of the $i$th particle and
$z_0$ is a complex constant related to the splitting of the zeros
due to anisotropy. This zero-splitting effect in the wavefunction
can be detected using the pair correlation function defined as\cite{yoshioka:2002}
\begin{eqnarray}
g(\mathbf{r})=\frac{L_{x}L_{y}}{N_{e}(N_e-1)}\langle 0|\sum_{i\neq
j}\delta(\mathbf{r}-(\mathbf{r}_i-\mathbf{r}_j))|0\rangle.
\label{paircorr}
\end{eqnarray}\\

In Fig. \ref{gr}(a), we plot the pair correlation functions along
$\hat{x}$ and $\hat{y}$ directions for the $\nu=1/3$ FQH state with
the interaction anisotropy $A_c=1.8$. The two curves are
distinguishable from their isotropic counterparts. We point out that
the curves behave asymptotically as $\alpha r^2$ in the limit of $r
\rightarrow 0$, with the prefactor $\alpha \propto |z_0|^4$. This is
entirely different from the isotropic Laughlin wavefunction, which
exhibits a $r^6$ asymptotic behavior in its pair correlation
function. The nonmonotonic behavior in $g(y)$ at small $r$ region,
which manifests itself more clearly at a larger $A_c$, is also consistent
with the zero-splitting scenarios.\cite{ciftja:2001,qiu:2012} We
point out that for a suitable deformed model wavefunction, there is
also an additional contribution to the Gaussian Landau level form
factor,\cite{read:2011,qiu:2012} which can be observable in disk
geometry with a boundary.

In Fig. \ref{gr}(b), we plot the square root of the prefactor
$\alpha$ at several small anisotropy.  The linear fit of
$\sqrt{\alpha}$ to $A_c$ is expected as $(A_c - 1)$ [or $(\sqrt{A_c}
- 1)$], which characterizes the perturbation away from the isotropic
point. However, the resulting nonzero intercept at $A_c = 1$
suggests that we may have overestimated the prefactor, possibly due
to the higher order contributions at small $A_c$.

\subsection{Hall-Smectic-like Phase in the Intermediate Interaction Anisotropy Regime}
\label{transition}

The explicit construction~\cite{qiu:2012} of the model
wavefunction by unimodular transformation on disk geometry suggests
that the geometrical description of the quantum Hall system accepts
the following deformation of the isotropic Laughlin state (i.e.,
$\gamma = 0$)
\begin{equation}
\Psi =
\prod_{i < j} z_{ij} \left [z_{ij}^2 + \frac{12 \gamma^*}{1 - \vert \gamma \vert^2 }  \right ]
e^{- \sum_i \gamma z_i^2/4} e^{- \sum_i \vert z_i \vert^2/4},
\end{equation}
where $z_{ij} = z_i - z_j$ and $\gamma$ characterizes the amount of
mixing between the guiding center creation and annihilation
operators in the unimodular transformation. Note that the model
wavefunction is expected to be valid for small $\gamma$. In the
present parametrization $\gamma = \sqrt{A_L^*} - 1$ is real. We can
postulate the breakdown criterion for the anisotropic Laughlin
liquid to be
\begin{equation}
\pi \frac{12 \gamma^*}{1 - \vert \gamma \vert^2 } = {2 \pi \over \nu},
\end{equation}
i.e., the area occupied by a set of three splitting zeros is the
average area per particle. This suggests that at the breakdown
$A_L^* \approx 2$ (i.e., $A_c \approx 3$ according to the estimation in
the subsection \ref{Laughlin state}), consistent with the onset of
the rapid decrease of the excitation gap.

In other words, the anisotropic Laughlin liquid is stable when  the
long-distance (i.e., at average particle spacing) behavior of the
Jastrow factor is still as $z_{ij}^{1/\nu}$. The collective
excitation of the liquid is the neutral magnetoroton excitations,
which becomes anisotropic. When the liquid phase breaks down, it
cannot sustain further anisotropy by the spatial deformation in the
roton spectrum.

\begin{figure}[t]
\centerline{\includegraphics [width=3.4 in] {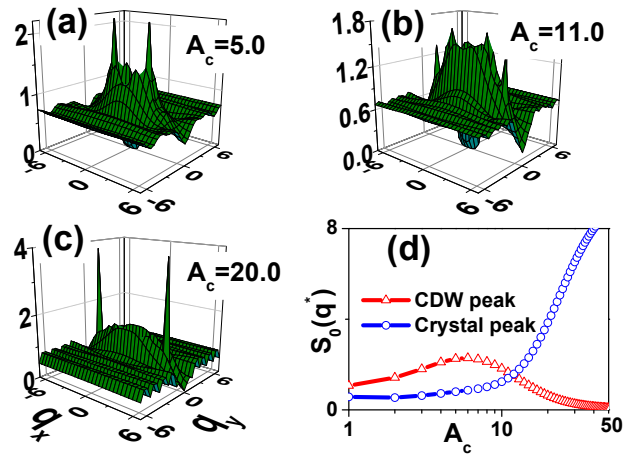}}
\caption{(Color online) (a)-(c): Structure factors for the ground
states of the $N_e=10$ system at $A_c=$ 5.0, 11.0, and 20.0,
respectively. Note that the locations of the twin peaks for $A_c =
5.0$ (CDW-like peaks) differ from those for $A_c = 20.0$ (Crystal
peaks). Both types of peaks coexist at $A_c=11.0$. (d) Peak values
of the CDW-like and Crystal peaks are plotted as a function of the
interaction anisotropy in a range of $1 < A_c < 50$.} \label{sqstar}
\end{figure}

One possible outcome of the system after this breakdown is that the mode
at the roton minimum goes softer,
developing some charge-density-wave (CDW) order. Due to the orientation effect of the anisotropy, this CDW is expected to be unidirectional (stripe-like) and the
characterizing sharp peaks in the structure factor are along the stretching direction. This is clearly visible in the structure
factor at $A_c = 5.0$ in Fig. \ref{sqstar}(a). The background in the
structure factor resembles that of an anisotropic Laughlin liquid,
but its peak value is significantly smaller than the two sharp peaks
along the $q_x$ axis. The CDW twin peaks correspond to a period in
real space, which can be roughly anticipated as the splitting of
zeros $\vert z_0 \vert = \sqrt{6}$ at the critical $A_L^* = 2$. The
peak value (subtracting background) in the structure factor suffices
as the order parameter. The plot in Fig. \ref{sqstar}(d) shows that
this CDW-like order parameter rules the system in the regime $2.0 <
A_c < 8.0$.

We term this phase, which breaks one-dimensional
translational symmetry, as a Hall-smectic-like phase since we speculate that it is related to the Hall smectic discussed
earlier in the context of liquid crystal phases in the FQH
system.\cite{fradkin:1999,macdonald:2000,barci:2002,musaelian:1996,balents:1996,fogler:2004} The rise of
the smectic phase softens the magnetoroton mode and appears to be
responsible for the reduction of the excitation gap for $2.0 < A_c <
8.0$ as shown in Fig.~\ref{Egap}. As discussed in
Ref.~[\onlinecite{balents:1996}] the transition from the Laughlin
liquid to the Hall smectic can be second order and its critical
behavior is in the $XY$ universality class.

Beyond $A_c=8.0$, the reverse trend in the excitation gap as a
function of $A_c$ indicates that the system is under the influence
of a distinct mechanism. This is evidently shown in the structure
factor plot of Fig. \ref{sqstar}(b) at $A_c=11.0$. Two additional
peaks along the $q_x$ axis are clearly visible with the different
wavevectors from the CDW-like twin peaks. These additional twin
peaks are corresponding to the unidirectional crystal order in the
quasi-1D repulsion limit that we will discuss in subsection
\ref{crystal}. The peak value of them is plotted as the crystal
order parameter in Fig. \ref{sqstar}(d). There we can see that the
crystal order parameter is continuously increasing with the
interaction anisotropy. The crossover for the competition with the
CDW-like order occurs around $A_c=13.0$. For larger anisotropy the
 crystal order dominates as illustrated in Fig. \ref{sqstar}(c) at
$A_c=20.0$, where only the crystal peaks remain. Thus, the Hall-smectic-like phase is found
unstable towards a 1D crystal.

\subsection{Quasi-1D Crystal in the Large Anisotropy Limit}
\label{crystal}

\begin{figure}[t]
\centerline{\includegraphics [width=3.4 in] {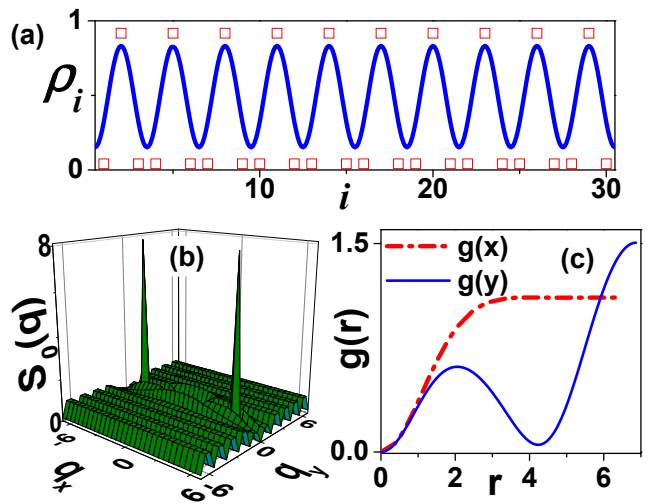}}
\caption{(Color online) For ground states of $N_e=10$ system at
$A_c=40.0$: (a) Average occupations (scattered squares) of the
Landau-level orbital at guiding centers and the charge density
(solid line) are plotted along the hard axis. A period of
$\lambda^*=L_x/N_e$ is clearly visible. The density values have been
shifted and exaggerated to emphasize a $\sim 2\%$ fluctuation over
the background. (b) Structure factor with sharp twin peaks in
$\hat{q_x}$ direction and periodic oscillation in $\hat{q_y}$
direction. (c) Pair correlation function is plotted along two
principal axes. The function $g(y)$ shows a oscillation with the
period $L_y/3$.} \label{rho}
\end{figure}

The crystal phase at the large anisotropy limit can be probed using
the charge distribution. In Fig. \ref{rho}(a), we plot the average
LL orbital occupations and the charge density along the hard axis at
$A_c=40.0$. The charge density appears smoother as an integral from local Gaussian wave packets over
orbital occupations and a $2\%$
fluctuation above the background can be observed in the exaggerated plot.
Both the charge occupation and density fluctuate along the hard axis
with the crystalline period $\lambda^*=L_x/N_e$. The maximum charge
occupation is close to unity as expected in the ultimate 1D crystal limit.
The 2D distribution of the charge density reveals that the system is
a unidirectional crystal with each electron spreading into a stripe
perpendicular to the hard axis. For basis states in the torus
geometry, the guiding-center coordinate along the hard axis is
coupled with the momentum along $q_y$ axis.\cite{yoshioka:2002}
Thus, the calculated ground states are expected to carry a period of
$2\pi/\lambda_y$ in the momentum space along the $q_y$ axis, where
$\lambda_y =L_y/3$. The structure factor plot in Fig. \ref{rho}(b)
demonstrates this characteristic order along $q_y$ axis. As its real
space counterpart, the pair correlation plot in Fig. \ref{rho}(c)
shows oscillation in $\hat{y}$ direction with a period $\lambda_y$.

The above results support that at filling $\nu=1/3$ the  ground
state of the system is a crystal in the large interaction anisotropy
limit and the system undergoes some transition from an
incompressible liquid to a solid as anisotropy increases. A similar
story has been discussed in an isotropic FQH system with extreme
geometry, such as in a thin torus or a cylinder
limit,\cite{rezayi:1994,lee:2004,bergholtz:2005} and in a recent
work \cite{wang:2011} on the graphene ribbon with flat bands. They
can be explained under the same principle in Ref.
~[\onlinecite{bergholtz:2005}] by sorting Hamiltonian. When the
interaction anisotropy increases, the repulsion-related diagonal
terms dominate, which has the similar effect as geometry on the
isotropic FQH and as the local orbital expansion on the flat-band
graphene ribbon. The low-energy physics is governed by the strong
repulsion so that the system tends to form crystal. At small
anisotropy, the hopping-related off-diagonal terms are comparable
and screen the repulsion, resulting in the liquid phase.

\subsection{Generalized Variational Laughlin State}\label{Laughlin state}

In the discussion above, we have seen that the isotropic Laughlin
wavefunction  is insufficient to fully capture the features of an
anisotropic FQH system. For such a system at the lowest LL filling
$\nu=1/q$, Haldane has suggested to use a family of Laughlin
states,\cite{haldane:2011} which is generally defined as the densest
zero-energy eigenstate of a projected two-body anisotropic
Hamiltonian:
\begin{eqnarray}
H_{v}(A_L)=\sum_{m<q}P_m(A_L).
\end{eqnarray}

For the fermion system with an odd denominator $q$,  $m$ are limited
to be odd. This Hamiltonian is a truncated summation over
anisotropic pair interactions:
\begin{eqnarray}
P_m(A_L)=\frac{1}{N_{\phi}} \sum_{\mathbf{q}} L_m(Q^2) e^{-Q^2/2}
\sum_{i<j}e^{i \mathbf{q} \cdot (\mathbf{R}_i-\mathbf{R}_j)}
\end{eqnarray}
for two particles with the relative angular momentum of $m\hbar$ in
the guiding-center coordinates. In the above expression, $L_m(x)$
are $m$th Laguerre polynomials and $Q(A_L)=\sqrt{q_x^2/A_L+A_L q_y^2}$,
which, like in Eq.~(\ref{metric}), defines a wavefunction metric
parameterized by $A_L$. The parameterized Laughlin states
$\Psi(A_L)$ satisfy
\begin{eqnarray}
P_m(A_L)|\Psi(A_L)\rangle=0, m<q.
\end{eqnarray}
The isotropic Laughlin wavefunction corresponds to the Laughlin
state with $A_L=1$. With this family of parameterized states, we are
able to variationally approximate the ground state of a FQH system
with anisotropic interaction. According to Haldane's proposal, if
the mass or orbital  metric (in our case, an identity matrix for
isotropic mass) is different from the interaction metric
(parameterized by $A_c$), the resulting variational state
$\Psi(A_L^*)$ should be described by a metric interpolating the mass
metric and the interaction metric, i.e., $1 < A_L^* < A_c$. This
intrinsic metric describes how correlated quasi-particles
effectively {\it feel} each other in a such anisotropic FQH system.

\begin{figure}[t]
\centerline{\includegraphics [width=3.0 in] {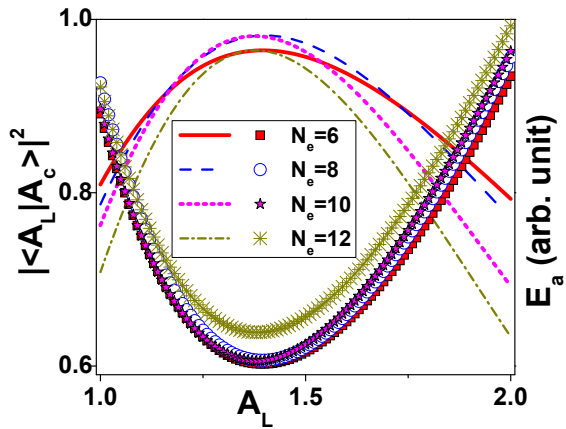}}
\caption{(Color online) Wavefunction square overlap (lines) and
expected value of Coulomb energy (symbols) as a function of the
variational parameter $A_L$ for $\nu=1/3$ FQH system at interaction
anisotropy $A_c=1.8$. The energy values have been shifted and
enlarged to emphasize that the location of the minimum coincides
with that of the largest overlap. The comparison of different sizes
of $N_e=6$, 8, 10, and 12 shows a weak size dependence only.}
\label{Ea}
\end{figure}

In Fig. \ref{Ea}, we study the anisotropic $\nu=1/3$ FQH system with
the Coulomb anisotropy $A_c=1.8$. The optimal Laughlin state
$\Psi(A_L^*)$ is obtained by tracing either the maximum of the
wavefunction overlap or the minimum of the expected Coulomb energy
\begin{eqnarray}
E_a(A_L)=\langle\Psi(A_L)|H_c(A_c)|\Psi(A_L)\rangle.
\end{eqnarray}
The optimal parameter is found at $A_L^* \sim 1.43$, which is weakly
size-dependent. This parameter is indeed an intermediate
value between unity and the Coulomb anisotropy as
expected.\cite{haldane:2011} It also agrees with the intrinsic
metrics through the analysis of the anisotropic structure factor in
the subsection \ref{liquid}. The overlaps between the optimal
Laughlin state and the exact ground states are larger than $90\%$ for
various system sizes, which supports the validity of the variational
state. We also note that the expected Coulomb energy quadratically
approaches its minimum, which suggests a linear approximation of the
anisotropic Laughlin state with $A_L$ in the liquid phase regime.

\begin{figure}[t]
\centerline{\includegraphics [width=3.0 in] {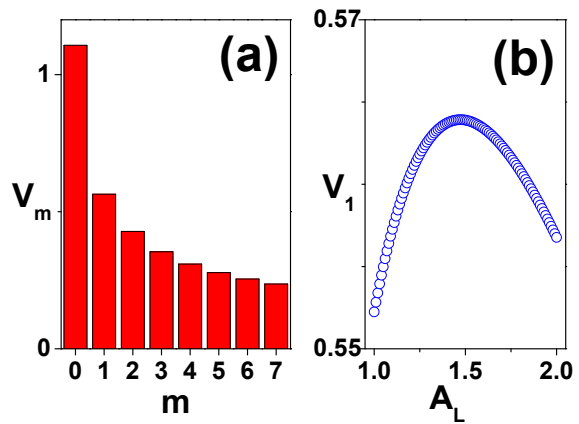}}
\caption{(Color online) For a Coulomb system with interaction
anisotropy $A_c=1.8$: (a) Effective pseudopotentials $V_m$ with a
given parameter $A_L=1.47$ is plotted against $m$. (b) The
pseudopotential $V_1$ is plotted as a function of $A_L$. The maximum
value occurs around $A_L=1.47$.} \label{pseudopotential}
\end{figure}

To gain a further understanding for the validity of this variational
approach, we approximately expand the Coulomb interaction in the
anisotropic pair interactions as:
\begin{eqnarray}
H_c(A_c)\approx \sum_{m} V_{m}(A_c,A_L)P_{m}(A_L),
\end{eqnarray}
where the average expansion coefficients $V_m$ define the effective
anisotropic pseudopotentials in a form of
\begin{eqnarray}
V_m=\int_{0}^{2\pi}d\theta\int_{0}^{\infty}dx\frac{L_m(x^2)e^{-F(\theta,x)/2}}{2\pi\sqrt{G(\theta)}}
\end{eqnarray}
with
\begin{eqnarray}
G(\theta)=(A_c/A_L)\cos^2\theta+(A_L/A_c)\sin^2\theta
\end{eqnarray}
and
\begin{eqnarray}
F(\theta,x)=x^2(1+A_L \cos^2\theta+\sin^2\theta/A_L).
\end{eqnarray}
An example of $V_m$ is plotted in Fig. \ref{pseudopotential}(a) with
$A_c=1.8$ and $A_L=1.47$. These pseudopotential parameters are found
positive and monotonously decrease with $m$, consistent with the
long range behavior of the Coulomb repulsion.

Given the Coulomb interaction anisotropy $A_c$, in principle we can
have a family of pseudopotential sets parameterized by $\{A_L\}$,
which are associated with different values. The set of $V_m$ with
the maximum pseudopotential values up to the $m$th order is most
promising for the truncation-based variational approach to work.
Thus, for the $\nu=1/3$ system, we could use the maximum of $V_1$ as
a criterion to examine the optimal parameter of the variational
Laughlin state, $A_L^*$. This {\it size-independent} condition
\begin{eqnarray}
V_1(A_L^*) \geq V_1({A_L}) \label{optimal}
\end{eqnarray}
serves as a semi-analytic estimation to the intrinsic geometry
parameter $A_L^*$ of the $\nu=1/3$ anisotropy system. As shown in
Fig. \ref{pseudopotential}(b), the estimated optimal parameter is
around $A_L^*=1.47$, matching the value found previously through the
finite-size calculation.

\begin{figure}[t]
\centerline{\includegraphics [width=3.4 in] {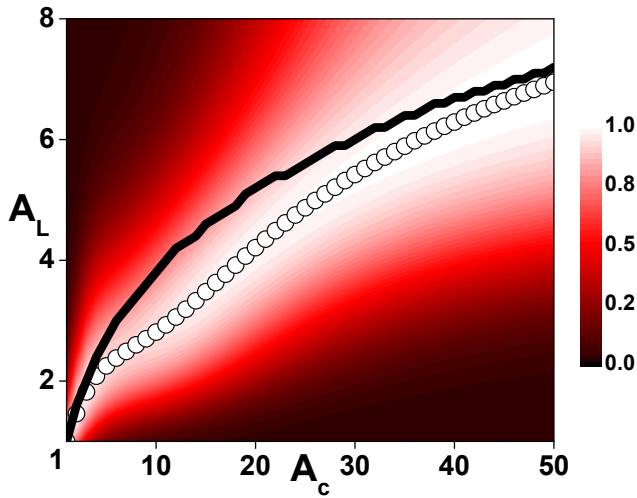}}
\caption{(Color online)  Contour plot of wavefunction square overlap
between the Coulomb ground state (parameterized by $A_c$) and the
generalized Laughlin state (parameterized by $A_L$) for the $N_e=10$
system. Each dot represents the maximum of the wavefunction overlap
for a given $A_c$. The solid curve stands for the estimated fit
$A_L^*$ as a function of $A_c$ using the maximum $V_1$ criterion.}
\label{ovlp}
\end{figure}

In the 2D contour plot of Fig. \ref{ovlp}, we show a comprehensive
map of the wavefunction overlap as a function of both parameters
$A_c$ and $A_L$. The isotropic Laughlin wavefunction has a
continuously decreasing overlap with the calculated FQH state when
the Coulomb anisotropy increases, indicating again its insufficiency
in describing the anisotropic FQH system. Instead, We note that at
the optimal parameters, the local maximum of the wavefunction
overlap is larger than $90\%$ in the {\it full} range of the Coulomb
anisotropy parameter ({\it even when the system is not in the liquid
phase}). This justifies that the family of variational Laughlin
states, which have larger wavefunction overlaps, could serve a
better description of the anisotropically interacting FQH system. We
also notice that in the vicinity of $A_c=1$, the optimal parameter
of the Laughlin state is linearly related to the interaction
anisotropy. As a comparison, we plot the estimated optimal
parameters $A_L^*$ from Eq. (\ref{optimal}) as a function of $A_c$. We
find they agree well with the finite-size results in the liquid and
crystal phases. But in the intermediate regime the results show some
deviation. This suggests a more careful handling beyond the simple
criterion of Eq. (\ref{optimal}) is needed for the intermediate
region.

\section{Summary}
\label{summary}

In conclusion, we have studied the effect of anisotropic Coulomb
interaction on the ground state of the 1/3 filling fractional
quantum Hall system. We find that at weak anisotropy the Laughlin
state remains to be a valid description, although the structure
factor and pair correlation function exhibit anisotropy. Our
calculations support the recent proposal of Haldane on the geometric
description of the fractional quantum Hall
state.~\cite{haldane:2011} In particular, the order-3 zeros in the
wavefunction split into three distinct zeros with a splitting distance
related to the anisotropy.~\cite{qiu:2012} We have compared the
ground state wavefunction of the anisotropic Coulomb interaction
with a family of single-parameter variational Laughlin states.  The
latter are obtained by deforming the projection (i.e., $V_1$ only)
Hamiltonian for the isotropic Laughlin state.  We have determined
the variational parameter by minimizing the variational ground state
energy or by maximizing the wavefunction overlap. In addition, we
also propose an effective analysis to estimate the optimal
variational parameter. The liquid phase breaks down when anisotropy
increases and a Hall-smectic-like order emerges. Finally, at strong interaction
anisotropy, which is more of theoretical interest, the ground state
exhibits a compressible one-dimensional crystal phase.
Interestingly, the ground state obtained by solving deformed
projection Hamiltonian remains to be a good description (overlap
greater than $90\%$) throughout the liquid to solid transition.

The anisotropy in Coulomb interaction provides a new route in
probing the intrinsic metric of the fractional quantum Hall state in
its geometrical description, as pointed out by
Haldane.~\cite{haldane:2011} In the present study we have revealed
that the intrinsic metric or the wavefunction anisotropy are indeed
different from the Coulomb metric or the dielectric tensor
anisotropy, although they appear to be linearly proportional to each
other in the vicinity of the isotropic Laughlin state. The linearity
we have found supports the proposal to use a single-parameter to
construct unimodularly deformed wavefunctions to describe the effect
of interaction anisotropy in disk geometry.~\cite{qiu:2012} In fact,
the family of deformed wavefunctions contain the same anisotropic
Jastrow factor in their relative coordinate
part,\cite{musaelian:1996} which explicitly splits the order-3 zeros
in the Laughlin liquid. Therefore, the emergence of the Hall smectic
phase, as analyzed in the effective field
theories,~\cite{balents:1996,fogler:2004} at larger anisotropy is
also a support of the geometrical description of FQH states.

Near the completion of this work, we note a very recent work
\cite{yang:2012} which discusses the anisotropic FQH system with
the anisotropic band mass. There, the persistent energy gap and anisotropic roton-minimum excitation at small anisotropy parameters are reported, consistent with our results in the liquid phase.

\section{Acknowledgments}

We thank Weiqiang Chen for discussions.  XW thanks Ruizhi Qiu
and Su Yi for a related collaboration on an anisotropic dipolar
interaction in ultracold fermion systems. This work was supported by
an RGC grant in Hong Kong, the National Basic Research Program of China
(973 Program) grant No. 2012CB927404, the National Natural Science
Foundation of China (NSFC) grant No. 11174246, and the DST India project
SR/S2/HEP-012/2009.

\end{document}